\documentclass[11pt]{article}
\usepackage[final]{acl}
\usepackage{times}
\usepackage{latexsym}
\usepackage{amsmath}
\usepackage[T1]{fontenc}
\usepackage{amsmath}
\usepackage{amssymb}
\usepackage[utf8]{inputenc}
\usepackage{url}
\usepackage{microtype}

\usepackage{inconsolata}
\usepackage{graphicx}

\title{Overview of the ClinicalSkillQA 2026 Shared Task on Continuous Perception and Procedural Reasoning in Clinical Skill Assessment}

\author{
\textbf{Xiyang Huang\textsuperscript{1,2\textsuperscript{*}}} \quad
\textbf{Renxiong Wei\textsuperscript{3\textsuperscript{*}}} \quad
\textbf{Yihuai Xu\textsuperscript{1,2}} \quad
\textbf{Zhiyuan Chen\textsuperscript{1,2}} \quad
\textbf{Keying Wu\textsuperscript{1,2}} \\
\textbf{Jiayi Xiang\textsuperscript{1,2}} \quad
\textbf{Buzhou Tang\textsuperscript{4}} \quad
\textbf{Yanqing Ye\textsuperscript{3}} \quad
\textbf{Jinyu Chen\textsuperscript{3}} \quad
\textbf{Cheng Zeng\textsuperscript{1}} \\
\textbf{Min Peng\textsuperscript{1,2,$\dagger$}} \quad
\textbf{Qianqian Xie\textsuperscript{1,2,$\dagger$}} \quad
\textbf{Sophia Ananiadou\textsuperscript{5}} \\
\\
\textsuperscript{1}School of Artificial Intelligence, Wuhan University \\
\textsuperscript{2}Center for Language and Information Research, Wuhan University \\
\textsuperscript{3}Zhongnan Hospital of Wuhan University \\
\textsuperscript{4}Harbin Institute of Technology, Shenzhen \quad
\textsuperscript{5}The University of Manchester \\
}

\begin{document}
\maketitle

\begin{abstract}
This paper presents an overview of the ClinicalSkillQA 2026 shared task, which was organized with the BioNLP Workshop at ACL 2026. The goal of this shared task is to evaluate continuous perception and procedural reasoning in clinical skill assessment by requiring systems to reconstruct the correct temporal order of shuffled clinical key frames and generate rationales grounded in clinical workflow knowledge. The benchmark contains 200 test-only instances sampled from clinical skill videos, covering three emergency-care procedures. Each instance is annotated with the ground-truth temporal order and an expert-verified rationale. A total of seven teams participated in the task, collectively making 90 submissions, with four teams providing system description papers. Systems are evaluated using Task Accuracy, Pairwise Accuracy, and BERTScore, which measure exact sequence reconstruction, local temporal consistency, and rationale quality, respectively. In this paper, we describe the task setup, dataset construction, and evaluation criteria. We further summarize the methodologies adopted by participating teams and present a comprehensive analysis of the submitted systems. The official results suggest that current models still struggle with continuous perception and procedural reasoning, especially when they must integrate visual evidence, temporal structure, and clinical workflow knowledge.
\end{abstract}

\section{Introduction}

Clinical skill assessment is a core component of medical education, where trainees are evaluated on whether their procedures follow standardized clinical workflows and safety-critical operational requirements~\citep{kogan2009tools, kogan2011opening}. 
In scenarios such as cardiopulmonary resuscitation, automated external defibrillator operation, and bag-mask ventilation, expert assessors review trainees' operations and judge whether each action is performed in the correct procedural context~\citep{harden1975assessment,regehr1998comparing,hodges1999osce}. 
This process requires more than recognizing isolated visual events: assessors must understand how clinical actions unfold over time, how earlier steps constrain later ones, and how a sequence of observations corresponds to a valid medical procedure. 
As multimodal large language models (MLLMs) are increasingly explored for medical education and clinical decision support, it becomes important to evaluate whether they can connect visual evidence with procedural medical knowledge and provide explanations that are meaningful to clinical experts.

Recent benchmarks have advanced the evaluation of multimodal reasoning in both general and medical domains~\citep{yue2024mmmu,zuo2025medxpertqa,yu2025medframeqa,xun2025rtv,wang2025continuousperceptionmattersdiagnosing}. 
However, many existing tasks focus on static image understanding~\citep{liu2024mmbench}, multi-image question answering~\citep{liu2024mibench}, or general-domain temporal reasoning~\citep{song2025burn}, leaving clinical procedural understanding underexplored. 
Medical benchmarks such as MEDFRAMEQA~\citep{yu2025medframeqa} require reasoning over multiple medical images, but the questions can often be answered by aggregating evidence from individual frames, without explicitly modeling the temporal dependencies among clinical actions. 
Similarly, continuous perception benchmarks such as CP-Bench~\citep{wang2025continuousperceptionmattersdiagnosing} evaluate whether models can integrate information across visual observations, but often simplify continuous perception into tasks such as counting, where the temporal order of observations may not directly affect the final answer. 
These settings are insufficient for clinical skill assessment, where the interpretation of an action is inherently conditioned on its position within a procedural workflow.

This limitation becomes especially important in clinical training scenarios. 
For example, a frame showing chest compressions cannot be fully interpreted without considering whether patient assessment, emergency response activation, and preparation steps have occurred beforehand. 
Likewise, a frame showing defibrillator use only becomes clinically meaningful when placed within the correct sequence of rhythm assessment, pad placement, safety checks, and shock delivery. 
Thus, evaluating MLLMs for clinical skill assessment requires tasks that go beyond recognizing what appears in each frame and instead test whether models can reconstruct temporally coherent and clinically plausible action progressions. 
It also requires explanation-based evaluation, since in medical education a model's answer is useful only when its reasoning can be inspected and aligned with clinical workflow knowledge.

To address this challenge, we organize the ClincalSkillQA shared task for evaluating continuous perception and procedural reasoning in clinical skill assessment. 
The task asks MLLMs to arrange shuffled key frames into the correct temporal order of clinical actions and to generate rationales explaining the predicted order. 
This formulation directly tests whether models can infer action progression from visual observations while grounding their decisions in clinical workflow knowledge. 
We provide 200 sets of shuffled key frames extracted from clinical skill videos covering three representative emergency-care procedures. 
Each sample is annotated with an expert-defined ground-truth ordering and a reference rationale. 
For evaluation, we use Task Accuracy to measure exact sequence reconstruction, Pairwise Accuracy to assess local temporal consistency between adjacent clinical actions~\citep{wang2025enact}, and BERTScore~\citep{zhang2019bertscore} to evaluate the semantic quality of generated rationales. 
ClinicalSkillQA provides a clinically grounded testbed for studying how MLLMs connect visual observations, temporal structure, and procedural medical knowledge in medical education scenarios.
\section{Task Description}
\label{sec:task}

ClinicalSkillQA formulates clinical skill understanding and continuous perception in clinical skill assessment as a multi-image temporal ordering task. Given a set of shuffled key frames extracted from a clinical skill video, a model is required to arrange the frames into a coherent sequence of clinical actions and provide a textual explanation for the resulting order. The task is designed to evaluate whether MLLMs can infer temporal continuity, procedural dependencies, and action-state transitions from fragmentary visual observations, rather than treating each frame as an independent image.

\subsection{Task Definition}
For each instance, the input consists of a set of shuffled key frames:
\[
\mathcal{I}=\{I_1, I_2, \ldots, I_n\},
\]
where each frame is associated with a unique identifier, such as A, B, C, or D. The frames are sampled from a continuous clinical skill video and randomly permuted before being presented to the model. The model is asked to output the correct temporal order of the frame identifiers and to generate an explanation that justifies this order.

The ground-truth answer is an ordered sequence:
\[
I^*=(I^*_1,I^*_2,\ldots,I^*_n),
\]
where $I^*_t$ denotes the frame identifier corresponding to the $t$-th step in the original clinical procedure. The model prediction is formulated as:
\[
\hat{y}=f_{\theta}(\mathcal{I}, \mathcal{P}),
\]
where $f_{\theta}$ denotes the evaluated model and $\mathcal{P}$ denotes the task instruction that specifies the required output format and asks the model to recover the temporal order of the shuffled frames with an explanation. The prediction $\hat{y}$ contains two components:
\[
\hat{y}=(\hat{I},\hat{R}),
\]
where $\hat{I}=(\hat{I}_1,\hat{I}_2,\ldots,\hat{I}_n)$ is the predicted temporal order of the provided frame identifiers, and $\hat{R}$ is the generated rationale explaining the temporal dependencies and procedural cues used to infer this order. A valid prediction should include each provided frame identifier exactly once. Predictions with missing, duplicated, or invalid identifiers are regarded as invalid outputs by the official evaluation script.

%\subsection{Clinical Scenarios and Data Source}

%ClinSkillQA is built on 200 sets of shuffled key frames extracted from three types of clinical skill videos. Each set represents a sequence of continuous clinical actions and is accompanied by expert-annotated ground-truth ordering and order rationales. The videos are derived from the SiMing-Bench\citep{huang2026siming} dataset, which contains medical student clinical procedure videos collected under standardized clinical skills assessment settings. The study was approved by the Institutional Review Board, and all data collection and processing followed relevant ethical guidelines.

\subsection{Evaluation Setting}

The official shared task evaluates both the predicted ordering and the generated rationale. For the ordering results, we use Task Accuracy and Pairwise Accuracy. Task Accuracy measures whether the predicted sequence exactly matches the ground-truth order. Pairwise Accuracy measures the fraction of correctly ordered adjacent frame pairs, providing a more fine-grained assessment of local temporal consistency.

For rationale evaluation, we use BERTScore to measure the semantic similarity between the generated rationale and the expert-annotated reference rationale. Together, these metrics assess whether a model can recover the correct temporal order while also producing explanations that are consistent with clinical workflows and expert reasoning.

\section{Dataset}
\label{sec:dataset}

ClinicalSkillQA is constructed from clinical skill assessment videos and consists of 200 test instances of shuffled key frames. Each instance corresponds to a continuous clinical action sequence and is associated with a hidden ground-truth temporal order and an order rationale. The dataset is designed as a test-only shared task benchmark to evaluate whether MLLMs can reconstruct coherent clinical procedures from fragmentary visual observations. As summarized in Table~\ref{tab:dataset_stats}, ClinicalSkillQA covers three clinical skill scenarios and uses shuffled key frames as input, with temporal order and rationale as evaluation targets.

\subsection{Data Source}

The videos are derived from the SiMing-Bench~\citep{huang2026siming} dataset, which contains medical student clinical procedure videos collected under standardized clinical skills assessment settings. We sample key frames from the videos to construct multi-image instances, where each instance represents a continuous clinical action sequence. This source provides realistic yet controlled procedural scenarios for evaluating continuous perception in medical contexts. Compared with independent medical image recognition, the sampled frames require models to infer temporal and procedural dependencies among clinically related actions, such as preparation, patient contact, device placement, action execution, and post-action checking. 

The study was approved by the Institutional Review Board. During data collection and processing, privacy protection procedures were applied to remove or mask personally identifiable information. The dataset is intended for research on multimodal clinical skill understanding and medical education, rather than for direct clinical decision-making.

\begin{table}[t]
\centering
\small
\setlength{\tabcolsep}{6pt}
\begin{tabular}{ll}
\hline
\textbf{Item} & \textbf{Description} \\
\hline
Dataset & ClinicalSkillQA \\
\# Instances & 200 \\
Scenarios & CPR, AED, BMV \\
Input & Shuffled key frames \\
\# Frames & 4 or 6 per instance \\
Target & Order \& rationale \\
Evaluation setting & Test-only benchmark \\
\hline
\end{tabular}
\caption{Dataset summary of ClinicalSkillQA. Each instance consists of 4 or 6 shuffled key frames sampled from a continuous clinical skill video. The hidden temporal order and expert-verified rationale are used for official evaluation.}
\label{tab:dataset_stats}
\end{table}

\subsection{Annotation}

Each instance is annotated with two types of supervision: the ground-truth temporal order and an order rationale. The temporal order is derived from the original chronological sequence of the sampled key frames. The rationale explains why the recovered order is clinically and procedurally correct.

We adopt a model-assisted and physician-verified annotation pipeline to improve both efficiency and consistency. Specifically, a multimodal model\footnote{We used Gemini 3 Pro for the first-stage rationale generation.} is first prompted with the shuffled key frames and the corresponding ground-truth order to generate an initial order rationale. 
The prompt asks the model to focus on visible action progression, state transitions, and clinically meaningful dependencies across frames. 
The generated rationale is then refined by a second multimodal model\footnote{We used GPT-5.2 for the second-stage rationale refinement.} to improve clarity, remove redundant descriptions, and ensure that the explanation is aligned with the visual evidence.

Finally, two physicians independently review the temporal order and the refined rationale. They correct clinically inaccurate statements, remove unsupported visual claims, and verify that the rationale reflects the actual procedural progression shown in the frames. Disagreements are resolved through discussion until consensus is reached. The final annotations are used only for official evaluation and are not released to participants during the shared task.

\section{Evaluation}

ClinicalSkillQA evaluates model performance from two perspectives: the correctness of the predicted temporal ordering and the quality of the generated rationale. 
Following the task definition, each prediction contains an ordering result $\hat{I}$ and an ordering rationale $\hat{R}$. 
We use Task Accuracy and Pairwise Accuracy to evaluate the predicted ordering, and BERTScore to evaluate the rationale.

\subsection{Task Accuracy}

Task Accuracy measures whether the model reconstructs the entire temporal sequence exactly. 
For the $i$-th instance, let the ground-truth ordering be
\[
I^{*(i)}=(I^{*(i)}_1,I^{*(i)}_2,\ldots,I^{*(i)}_{n_i}),
\]
and the predicted ordering be $\hat{I}^{(i)}$. 
The prediction is counted as correct only when the two sequences are identical:
\[
\operatorname{Acc}_i =
\mathbb{I}\left(\hat{I}^{(i)} = I^{*(i)}\right),
\]
where $\mathbb{I}(\cdot)$ is the indicator function. 
Given $N$ test instances, Task Accuracy is computed as
\[
\operatorname{TaskAcc}
=
\frac{1}{N}
\sum_{i=1}^{N}
\mathbb{I}
\left(
\hat{I}^{(i)} = I^{*(i)}
\right).
\]

This metric is strict: any misplaced, missing, or duplicated frame identifier causes the instance to be marked as incorrect.

\subsection{Pairwise Accuracy}

Task Accuracy does not reflect partial temporal correctness.
Therefore, we further report Pairwise Accuracy, which measures whether the model preserves the correct relative order of adjacent frame pairs in the ground-truth sequence.

For the $i$-th instance, let $\operatorname{pos}_{\hat{I}^{(i)}}(x)$ denote the position of frame identifier $x$ in the predicted ordering.
For each adjacent ground-truth pair $(I^{*(i)}_t, I^{*(i)}_{t+1})$, we define
\[
c_{i,t} =
\mathbb{I}\left(
\operatorname{pos}_{\hat{I}^{(i)}}(I^{*(i)}_t)
<
\operatorname{pos}_{\hat{I}^{(i)}}(I^{*(i)}_{t+1})
\right).
\]
If a required frame identifier is missing, the corresponding pair is treated as incorrect.

We compute Pairwise Accuracy by aggregating all adjacent ground-truth pairs across the evaluation set:
\[
\operatorname{PairAcc}
=
\frac{
\sum_{i=1}^{N}
\sum_{t=1}^{n_i-1}
c_{i,t}
}{
\sum_{i=1}^{N}(n_i-1)
}.
\]

This metric only checks whether each adjacent ground-truth pair appears in the correct relative order in the prediction; the two frames do not need to remain adjacent in the predicted sequence.
Pairwise Accuracy provides partial credit for recovering local temporal transitions, even when the full sequence is not perfectly reconstructed.

\subsection{BERTScore for Rationale Evaluation}

We also evaluate the generated ordering rationale $\hat{R}$ against the expert-verified reference rationale $R^*$. 
BERTScore is used because it measures semantic similarity based on contextual embeddings rather than exact lexical overlap. 
This is suitable for ClinicalSkillQA, where different explanations may use different wording but still describe the same clinical operation, procedural stage, visual evidence, and temporal transition. 
We report the average BERTScore F1 over all test instances.

\subsection{Overall Score}

The official leaderboard score aggregates temporal ordering performance and rationale quality.
We first define the ordering score as
\[
S_{\mathrm{ord}}
=
\alpha \cdot \operatorname{TaskAcc}
+
(1-\alpha) \cdot \operatorname{PairAcc},
\]
where $\alpha$ controls the relative importance of exact sequence reconstruction and partial temporal consistency.

The final score is then computed as
\[
S_{\mathrm{final}}
=
100 \cdot
\left[
\beta \cdot S_{\mathrm{ord}}
+
(1-\beta) \cdot \operatorname{BERTScore}_{F1}
\right],
\]
where $\beta$ balances the ordering result and the generated rationale.
In our official evaluation, we set $\alpha=0.7$ and $\beta=0.8$.
This choice assigns higher importance to exact sequence reconstruction while still rewarding models for preserving local temporal transitions and generating semantically meaningful rationales.
The factor $100$ scales the score to a more readable range.

\section{Results}
\label{sec:results}

We used the CodaBench platform\footnote{\url{https://www.codabench.org/competitions/14884/\#/pages-tab}} to manage the submission and evaluation process for the shared task. Overall, seven teams participated and made a total of 90 submissions.

Table~\ref{tab:results} reports the official results\footnote{Component metrics are rounded to two decimal places, while overall scores are computed using unrounded official metric values.} on the ClinicalSkillQA test set.
Teams are ranked by the overall score, which jointly considers exact temporal ordering, local temporal consistency, and rationale quality.

\begin{table*}[t]
\centering
\small
\begin{tabular}{lcccc}
\hline
\textbf{Team} & \textbf{Overall Score} & \textbf{Task Accuracy} & \textbf{Pairwise Accuracy} & \textbf{BERTScore F1} \\
\hline
ZZUNLP & 71.43 & 0.63 & 0.86 & 0.79 \\
qqpprun & 63.88 & 0.58 & 0.84 & 0.55 \\
FBK & 59.45 & 0.47 & 0.72 & 0.76 \\
nance & 56.62 & 0.41 & 0.75 & 0.79 \\
VerbaNexAI & 37.96 & 0.17 & 0.60 & 0.71 \\
DLNLP & 30.44 & 0.10 & 0.55 & 0.57 \\
zhekun & 28.71 & 0.03 & 0.51 & 0.74 \\
\hline
\end{tabular}
\caption{
Official results on the ClinicalSkillQA test set.
Teams are ranked by the overall score.
Pairwise Accuracy denotes micro-averaged pairwise accuracy.
}
\label{tab:results}
\end{table*}

Overall, ZZUNLP achieves the best performance, with an overall score of 71.43.
It also obtains the highest Task Accuracy and Pairwise Accuracy, reaching 0.63 and 0.86, respectively, and achieves a strong BERTScore F1 of 0.79.
The second-ranked system, qqpprun, obtains an overall score of 63.88, with Task Accuracy and Pairwise Accuracy close to those of ZZUNLP.
This indicates that the leading systems are relatively effective at reconstructing temporal sequences and preserving local temporal transitions.

The middle-ranked systems, including FBK and nance, achieve overall scores of 59.45 and 56.62, respectively.
Although their ordering metrics are lower than those of the top two systems, they obtain strong BERTScore F1 scores.
In particular, nance reaches a BERTScore F1 of 0.79, matching that of ZZUNLP.
This suggests that rationale generation can remain competitive even when exact temporal reconstruction is less accurate.

The remaining systems obtain substantially lower overall scores, mainly due to limited Task Accuracy.
For example, VerbaNexAI, DLNLP, and zhekun achieve Task Accuracy scores of 0.17, 0.10, and 0.03, respectively, while their Pairwise Accuracy scores remain higher.
This gap indicates that these systems can recover partial local temporal relations but often fail to reconstruct the complete sequence exactly.
Notably, zhekun obtains a relatively high BERTScore F1 of 0.74 despite its low ordering accuracy, further showing that rationale similarity does not necessarily imply correct temporal ordering.

Across all teams, Pairwise Accuracy is consistently higher than Task Accuracy.
This trend shows that preserving local temporal relations is generally easier than reconstructing the full shuffled sequence exactly.
Meanwhile, BERTScore F1 is relatively high for several systems even when their Task Accuracy remains limited.
This suggests that rationale quality and ordering correctness capture complementary aspects of model performance: a system may generate clinically plausible explanations without fully recovering the correct temporal order.
Overall, these results demonstrate that ClinicalSkillQA remains challenging for current systems, especially in requiring models to jointly perform fine-grained temporal reconstruction and clinically grounded rationale generation.

\section{System Descriptions}

This section summarizes the submitted systems for the ClinicalSkillQA shared task. Out of the seven participating teams, four teams submitted system papers. Overall, participating teams explored different strategies for reconstructing the temporal order of clinical keyframes and generating corresponding rationales. The submitted methods mainly relied on multimodal large language models, with different designs for visual evidence extraction, procedural reasoning, and explanation generation.

\paragraph{ZZUNLP.}
ZZUNLP proposed Perceive-and-Plan, a two-stage in-context learning framework for clinical skill keyframe reordering. 
The system first performs structured visual perception by generating compact clinical descriptions for each frame, enhanced with saliency-guided picture-in-picture inputs that zoom in on critical head/airway and chest/hand regions. 
It then conducts temporal planning in a fresh conversation, using BLS procedural rules and visual anchors to verify the predicted chronological order. 
Without parameter updates, the system achieved an overall score of 71.43, with 0.63 Task Accuracy, 0.86 Pairwise Accuracy, and 0.79 BERTScore F1. 
The results suggest that separating visual perception from temporal reasoning improves local ordering reliability, while the remaining gap between exact sequence accuracy and pairwise accuracy indicates that long-range permutation errors and subtle state-transition ambiguities remain challenging.

\paragraph{FBK.}
FBK adopted a training-free prompting framework based on Qwen3-VL-32B-Instruct for clinical keyframe ordering. 
Their system examined several prompt formulations, random filename renaming to reduce alphabetical-order bias, few-shot demonstrations, and keypoint-derived predicates extracted by YOLOv8 pose estimation. 
The strongest setting used randomly renamed frames and full human-pose predicates without few-shot examples, achieving an overall score of 59.45, with 0.47 Task Accuracy, 0.72 Pairwise Accuracy, and 0.76 BERTScore F1. 
Their analysis shows that simple prompting was more effective than more complex intermediate reasoning prompts, while filename renaming consistently improved performance by reducing spurious textual biases. 
The largest gain came from keypoint-derived predicates, suggesting that explicit posture and body-motion cues provide useful temporal evidence for reconstructing clinical action sequences.

\paragraph{VerbaNexAI Lab.}
VerbaNexAI Lab proposed a zero-shot two-stage framework based on Qwen2-VL-2B-Instruct. 
The system first extracts frame-level visual evidence and maps it into structured procedural states using deterministic rules. 
It then assigns each frame an ordinal stage score and sorts frames accordingly, with PCA over multimodal embeddings used to break ties. 
The final prediction includes both the ordered frame sequence and an evidence-based rationale. 
On the official test set, the system achieved an overall score of 37.96, with 0.17 Task Accuracy, 0.60 Pairwise Accuracy, and 0.71 BERTScore F1. 
These results indicate that the method produced complete and semantically reasonable rationales, but still struggled with exact temporal ordering, especially when adjacent procedural states were visually similar.

\paragraph{DLNLP.} DLNLP proposed EvidenceFlow, a structured zero-shot framework built on Qwen2.5-VL-7B-Instruct for clinical keyframe ordering. 
The method decomposes the task into global overview, local evidence modeling, and ordering decision, and explores two variants: EvidenceFlow-M, which relies on model-led reasoning, and EvidenceFlow-R, which introduces rule-guided correction with explicit evidence constraints. 
On the official test set, the submitted EvidenceFlow-R variant achieved an overall score of 30.44, with 0.10 Task Accuracy, 0.55 Pairwise Accuracy, and 0.57 BERTScore F1. 
In their system paper, DLNLP further reports that EvidenceFlow-M achieved a higher BERTScore F1 of 0.74, whereas EvidenceFlow-R performed better on ordering-related metrics. 
These results suggest that explicit rule constraints improve ordering stability, whereas more flexible model-led reasoning may produce semantically richer explanations. 
The main remaining bottleneck is the stable integration of subtle cross-image temporal cues into a globally consistent clinical sequence.

\section{Conclusion}

We introduced ClinicalSkillQA, a BioNLP shared task for evaluating continuous perception and procedural reasoning in clinical skill assessment. Given shuffled key frames from clinical skill videos, systems are required to recover the correct temporal order and generate rationales explaining their predictions. The benchmark contains 200 test-only instances covering three emergency-care procedures, with annotations consisting of the ground-truth temporal order and expert-verified rationales. Official results show that current multimodal systems can capture some local temporal relations and produce plausible explanations, but still face clear challenges in exact full-sequence reconstruction. These findings suggest that clinical procedural reasoning remains difficult for existing multimodal large language models, particularly when visual evidence, temporal dependencies, and clinical workflow knowledge must be jointly integrated. ClinicalSkillQA provides a clinically grounded testbed for future research on reliable multimodal reasoning in medical education.

\section{Limitations}

ClinicalSkillQA has several limitations. First, the benchmark currently includes 200 test-only instances from three emergency-care procedures, which may not fully represent the diversity of clinical skills, medical specialties, and assessment environments. Expanding the dataset to more procedures and clinical contexts would improve its coverage and generalizability.

Second, the task is based on shuffled key frames rather than full videos. This formulation enables controlled evaluation of temporal ordering and procedural reasoning, but it does not fully capture motion dynamics, action duration, or fine-grained temporal transitions. Therefore, the benchmark primarily evaluates whether models can infer procedural order from representative visual observations, rather than their complete ability to understand continuous clinical videos.

Third, rationale quality is evaluated using BERTScore F1. Although this provides an automatic measure of semantic similarity, it may not fully reflect clinical correctness, visual grounding, or educational usefulness. Future work could introduce clinically informed rationale metrics or expert-based evaluation.

Finally, ClinicalSkillQA is designed as a test-only shared task benchmark. This setting supports fair evaluation of submitted systems, but it does not examine the effects of supervised training or domain adaptation. Future versions may include development data, additional procedures, and full-video assessment settings.


\bibliography{custom}

\appendix

%\section{Example Appendix}
%\label{sec:appendix}

%This is an appendix.

\end{document}